\documentclass{aastex63}

\usepackage{CJK}
\usepackage{amsmath}
\usepackage{textcomp}
\usepackage{amssymb}
\usepackage{tikz}
\usetikzlibrary{arrows}
 \usepackage{booktabs}
 \usepackage{multirow}
\usepackage{tabularx}
\usepackage{booktabs,multirow}

\definecolor{DarkGreen}{rgb}{0.0,0.4,0.0}  
\newcommand{\B}[1]{{\color{blue}{#1}}}
\newcommand{\R}[1]{{\color{red}{#1}}}
\newcommand{\G}[1]{{\color{DarkGreen}{#1}}}
\newcommand{\M}[1]{{\color{magenta}{#1}}}
\newcommand{\C}[1]{{\color{cyan}{#1}}}
\usepackage[normalem]{ulem}
\usepackage{soul}
\usepackage{cancel}

\submitjournal{ApJ}

\shorttitle{two thermal components}
\shortauthors{Zhou et al.}

\submitjournal{APJ}

\begin{document}
\begin{CJK*}{UTF8}{gbsn}
\title{Resolving Two Distinct Thermal X-ray Components in A compound Solar Flare}

\correspondingauthor{Zhenjun Zhou,Rui Liu}
\email{zhouzhj7@mail.sysu.edu.cn, rliu@ustc.edu.cn}

\author[0000-0001-7276-3208]{Zhenjun Zhou(周振军)}
\affiliation{Planetary Environmental and Astrobiological Research Laboratory (PEARL), School of Atmospheric Sciences, Sun Yat-sen University, Zhuhai, China}

\affiliation{Key Laboratory of Lunar and Deep Space Exploration, National Astronomical Observatories, Chinese Academy of Sciences, Beijing 100101, China}
\affiliation{CAS Center for Excellence in Comparative Planetology, China}

\author[0000-0003-4618-4979]{Rui Liu}
\affiliation{CAS Key Laboratory of Geospace Environment, Department of Geophysics and Planetary Sciences, University of Science and Technology of China, Hefei, Anhui 230026, China}

\author[0000-0001-5975-2651]{Jianqing Sun}
\affiliation{School of Astronomy and Space Science, Nanjing University, Nanjing 210023, China}
\author[0000-0003-0951-2486]{Jie Zhang}
\affiliation{Department of Physics and Astronomy, George Mason University, Fairfax, VA 22030, USA}
\author[0000-0002-4978-4972]{Mingde Ding}
\affiliation{School of Astronomy and Space Science, Nanjing University, Nanjing 210023, China}
\author[0000-0003-2837-7136]{Xin Cheng}
\affiliation{School of Astronomy and Space Science, Nanjing University, Nanjing 210023, China}
\author[0000-0002-8887-3919]{Yuming Wang}
\affiliation{CAS Key Laboratory of Geospace Environment, Department of Geophysics and Planetary Sciences, University of Science and Technology of China, Hefei, Anhui 230026, China}
\author{Xiaoyu Yu}
\affiliation{Planetary Environmental and Astrobiological Research Laboratory (PEARL), School of Atmospheric Sciences, Sun Yat-sen University, Zhuhai, China}
\author[0000-0001-6804-848X]{Lijuan Liu}
\affiliation{Planetary Environmental and Astrobiological Research Laboratory (PEARL), School of Atmospheric Sciences, Sun Yat-sen University, Zhuhai, China}
\author[0000-0002-4721-8184]{Jun Cui}
\affiliation{Planetary Environmental and Astrobiological Research Laboratory (PEARL), School of Atmospheric Sciences, Sun Yat-sen University, Zhuhai, China}

\begin{abstract}
X-ray emission provides the most direct diagnostics of the energy-release process in solar flares. Occasionally, a superhot X-ray source is found to be above hot flare loops of $\sim\,$10 MK temperature. While the origin of the superhot plasma is still elusive, it has conjured up an intriguing image of in-situ plasma heating near the reconnection site high above the flare loops, in contrast to the conventional picture of chromospheric evaporation. Here we investigate an extremely long-duration solar flare, in which EUV images show two distinct flare loop systems that appear successively along an $\Gamma$-shaped polarity inversion line (PIL). When both flare loop systems are present, the HXR spectrum is found to be well fitted by combining a hot component ($T_{e}\sim$12 MK) and a superhot component ($T_{e}\sim$30 MK). Associated with a fast CME, the superhot X-ray source is located at top of the flare arcade that appears earlier, straddling and extending along the long `arm' of the $\Gamma$-shaped PIL. Associated with a slow CME, the hot X-ray source is located at the top of the flare arcade that appears later and sits astride the short `arm' of the $\Gamma$-shaped PIL. Aided by observations from a different viewing angle, we are able to verify that the superhot X-ray source is above the hot one in projection, but the two sources belong to different flare loop systems. Thus, this case study provides a stereoscopic observation explaining the co-existence of superhot and hot X-ray emitting plasmas in solar flares.

\end{abstract}
\emph{Online-only material}: animations, color figures
\section{Introduction}
The creation of hot (10--20 MK) plasma in the corona is a prominent feature in nearly all solar flares.
Plasma at such high temperatures emits soft X-rays (SXRs) from both 
bremsstrahlung and resonant lines of highly ionized atoms - primarily from Fe XXIV and XXV \citep{Korneev1979,Caspi2010}. 
During the flare impulsive phase, the light curve of SXR flux often shows a tendency to resemble  that of the time integral of hard X-ray (HXR) flux. This empirical relationship, which is also known as the Neupert effect \citep{Dennis1993}, has provided evidence for the thick-target model \citep{Brown1971} , in which the HXR emission is produced by the bremsstrahlung of energetic electrons as they are instantly thermalized in the dense chromosphere, presumably at the footpoints of newly reconnected field lines, which heats up the local chromospheric plasma to temperatures in excess of 10 MK;
the overpressure of the over-heated chromosphere propels hot plasma upward into the corona along the same field lines, forming X-ray emitting flare loops \citep{Antonucci1984,Fisher1985,Allred2005,Allred2015}. This upward flow is conventionally termed as chromospheric evaporation \citep{Antiochos1978,Cheng2019}, which is often thought to be the ubiquitous source of the $\sim\,$10--20 MK plasma observed in nearly all flares, whose temperatures are consistent with those derived from 
SXRs of the \emph{Geostationary Operational Environmental
Satellite} (\emph{GOES}).  
Termed loop-top source in the literature, an HXR thermal component with temperatures of 10--20 MK is often detected at the top of the SXR flare loops with indirect imaging methods \citep{Liu2013,Sun2014,Sun2016}.\par

In some flares, besides the ubiquitous loop top hot component (10--20 MK), a spatially distinct superhot ($T_{e}\geqslant$ 30 MK) thermal component has been reported. 
This superhot plasma is first unveiled with high-resolution HXR spectroscopy \citep{Lin1981}, which is characterized by a steeply-falling spectrum resembling that of $\sim\,$34 MK plasma. Continuum and Fe XXVI line observations showed such a superhot component generally exists in GOES X-class flares \citep{Tanaka1987,Pike1996}.
Due to the limited spatially resolved observations of high-temperature passbands,
the source region of this superhot component ($T_{e}\geqslant$ 30 MK) remains elusive. Case studies offer a glimpse of location of the superhot source region.
With the aid of direct SXR images by Yohkoh \citep{Ogawara1991}, \citet{Nitta1997} reported a flare with the superhot component consisting of two separate loop structures, with the dominant HXR flux from an extended structure away from the bright SXR loop. With the aid of HXR imaging and spectroscopy implemented by \emph{Reuven Ramaty High Energy Solar Spectroscopic Imager \citep[\emph{RHESSI};][]{Smith2002,Lin2002}, \citet{Caspi2010})}, found that, in a GOES X4.8 flare on 2002 July 23, the superhot plasma is located distinctly above the flare loop top containing the conventional 10--20 MK plasma, peaks simultaneously as the non-thermal HXRs, and exists even during the pre-impulsive phase with negligible footpoints. These observations suggest that the super-hot plasma is in-situ heated, i.e., more directly related to the accelerated non-thermal electrons and hence to the reconnection process than the cooler flare plasma, which is due to the traditional picture of chromospheric evaporation \citep{Caspi2010}. Employing HXR spectroscopy to derive the temperature of overall flare plasma, \citet{Caspi2014} found a strong correlation between the maximal temperature and the flare GOES class in 37 M-class-and-above flares. But ``super hot''  temperatures exceeding 30 MK are found almost exclusively in X-class flares. It is unclear, however, whether these flares contain both super-hot and hot components like the prototypical super-hot flares reported before \citep[e.g.,][]{Lin1981,Nitta1997,Caspi2010}.
Nevertheless, our knowledge about the superhot component, including its spatio-temporal relationship to the energy release and transport processes that are active within most flares, is still scarce.

Moreover, some flares have a SXR light curve containing two or more peaks, which are as close as minutes apart. These multi-peaks are often associated with two or more closely connected magnetic structures erupting consecutively within a short time interval. Such a flare is also termed a compound eruption \citep{Woodgate1984,Dhakal2018}.

In this paper, we investigate a compound flare that lasted for an extremely long duration and proceeded sequentially in space along a curved PIL. 
The long duration and the optimal projection provide us an excellent opportunity to analyze the thermodynamic evolution in both time and space of the flare. 
The flare is well observed close to the limb by 
\emph{Atmospheric Imaging Assembly} (\emph{AIA}, \citealt{Lemen2012}) filtergram and \emph{Helioseismic and Magnetic Imager} (\emph{HMI}, \citealt{Schou2012}) onboard \emph{Solar Dynamics Observatory} (\emph{SDO}, \citealt{Pesnell2012}),  \emph{RHESSI}, and \emph{GOES}.
Meanwhile, this flare is also well observed from an vantage point 
by Extreme Ultraviolet Imager (\emph{EUVI}, \citealt{Howard2008}) telescope onboard 
Solar TErrestrial RElations Observatory A (\emph{STEREO-A}, \citealt{Kaiser2008}),
which observed the flare as an ``on-disk'' event, with a spatial resolution of 1.6$\arcsec$ and a cadence of 10 minutes.

These multi-wavelength and dual-perspective observations enable us to conduct a comprehensive study of  the involved thermal structures and their dynamic evolution in this flare. The paper is organized as follows. We present the observations in Sect.~\ref{sec:Obs}. The data analysis and results are described in Sect.~\ref{sec:Analysis}, followed by a discussion and conclusions in Sect.~\ref{sec:Conclusions}.

\section{Observations} \label{sec:Obs}

On 2012 July 17, a GOES class M1.7 flare occurred near the southwestern limb of the Sun. This flare began at $\sim$12:24 UT and took $\sim$5 hours long to reach the SXR peak at $\sim$17:15 UT and then took over 7 hours to reach the pre-flare level, making the whole duration longer than 12 hours (Figure~\ref{f1}(a)). In contrast, a typical flare lasts from a few minutes to tens of minutes, and a long-duration flare lasts for hours, also known as long-duration-event (LDE) flare \citep{Sheeley1983,Webb1987}. It was termed `the slowest flare' by Sam Freeland and Hugh Hudson\footnote[1]{http://sprg.ssl.berkeley.edu/$\sim$tohban/wiki/index.php/The\underline{ }Slowest\underline{ }Flare}. 
In addition, we also measure the time derivative of GOES SXR flux \mbox{df/dt} during the solar flare (see red line in Figure~\ref{f1}(a)), which can be used as a proxy for the HXR flux according to the Neupert effect \citep{Dennis1993}.
It should be noted that RHESSI HXR emission (usually defined as X-ray emission above $\sim$20 keV) is not favored for this extended flare because of the frequent gaps and low count rates.
According to the three peaks in the time derivative of GOES SXR (red profile in Figure~\ref{f1}(a)),
the rising phase of the flare can be further divided into three episodes: Episode I, from $\sim$12:24 UT to $\sim$14:05 UT; Episode II, from $\sim$14:05 UT to $\sim$16:10 UT; and Episode III, from $\sim$16:10 UT to $\sim$17:35 UT. The first and last episodes were associated with two coronal mass ejections (CMEs) at $\sim$ 13:48 UT and $\sim$17:00 UT, respectively. Both CMEs launched towards the southwest (manifested by the central position angle (CPA), which is measured counterclockwise from the projection of the Sun's north pole of the broadside CMEs). The first was a fast CME with a velocity of 958 km/s while the latter was a slow CME with a velocity of 395 km/s (See the CME height-time plots in Figure~\ref{f1}(b)), implying the whole process was composed of at least two different loop systems instead of a superposition of an extended sequence of similar loops along the PIL. \par 

This extremely long-duration flare also spanned a large area in space,  approximately 250$\arcsec$ along the south-north direction in the NOAA active region (AR) 11520. The AR is characterized by a major sunspot of positive polarity surrounded by diffuse magnetic flux of negative polarities. As a result, the flaring PIL takes a $\Gamma$ shape, with the long `arm' in the N-S orientation and the short arm in the E-W orientation (Figure~\ref{f2}(e--g)).  
To investigate the spatial locations of the thermal components, we reconstruct the RHESSI X-ray sources in the energy bands of 6-25 keV, the
integration time of the images is 40s. We use the standard
image reconstruction CLEAN algorithm. 
The CLEAN method is an iterative algorithm \citep{Hurford2002}. It is widely employed in X-ray image reconstruction due to the excellent record of bringing out the X-ray image morphology. It is basically a process of ``deconvolution'' of the back-projected image using the point spread function (PSFs).
Detectors 3-9 are used, but without detector 4 because it is excessively noisy during this flare. For all other parameters, the defaults are used. \par

In different episodes, the flare showed distinct emission structures at different locations. In the first episode, there appeared a group of sheared loops in the AIA 131{\AA} passband in the southernmost part of the active region (Figure~\ref{f2}(a) and accompanying animation). These loops were relatively low in height with one compact footpoint patch rooting in the positive magnetic polarity (northwestern part) and one extended footpoint patch rooting in the negative magnetic polarity (southeastern part), where the corresponding brightenings were clearly observed in AIA 1600{\AA} passband (Figure~\ref{f2}(e)). Above the loops seen in EUV, there existed a 6-25 keV X-ray loop-top source, implying that these EUV loops were likely hot post-flare loops produced by magnetic reconnection in the corona. During the one and half hours long evolution of the first episode, the morphology and the height of the EUV loops did not show significant changes. However, the SXR emission intensity kept increasing. Besides, the flare ribbon in the negative polarity showed a considerable separation movement from the PIL and a northward expansion (Figure~\ref{f2}(f)). Above the stationary and low-lying EUV loops, there existed a faint large-scale loop-like structure visible in 131 {\AA} passband from 13:10 UT, which slowly rose for about 30 minutes and quickly erupted after 13:43 UT. This eruption resulted in a large CME seen in coronagraph images (black plus symbols in Figure~\ref{f1}(b)). \par

Following this quick eruption, the flare evolved into the second episode and showed a sequential evolution in space along the PIL from south to north (Figure~\ref{f2}(b) and (f)). The loop-top X-ray source in RHESSI became highly extended in the N-S direction. In addition, the flare ribbon in the negative polarity region quickly expanded northward in association with the sequential formation of the post-flare loop arcade seen in the 131 {\AA} passband. The flare ribbon initially developed in parallel with the PIL, then showed certain separation perpendicular to the PIL (Figure~\ref{f2}(f)), probably due to the ascent of the magnetic reconnection site. The second CME launched at around 16:20 UT (red plus symbols in Figure~\ref{f1}(b)). After that, a new group of post-flare loops became visible at $\sim$16:28 UT (Figure~\ref{f2}(c)), transiting into the third episode of the rising phase. \par

During the third episode, the flare arcade develops along the E-W oriented PIL segment toward the limb (Figure~\ref{f2}(d) and (g)). Thus, it became more difficult to observe the flare evolution along the PIL due to projection effects. Incorporating the STEREO observation (Figure~\ref{f5}(b) and accompanying animation), one can see that the increased brightness mostly came from a compact region at the northernmost part. \par

\section{Analyses and Results} \label{sec:Analysis}
The analysis of thermal properties of flare regions can help us infer where the energy is released from magnetic reconnection. Through tracking the evolution of thermal sources, we can also infer how the energy is transported from one place to other places in the flare region. The nature of the slow evolution and long duration of the flare studies here provides us an excellent opportunity to deduce a clear picture of how thermal plasmas evolve after being heated by magnetic reconnection. Moreover, analyzing the sequential evolution of the thermal sources  along the PIL helps improve our understanding of three-dimensional aspects of the flare process.\par

We derive the thermal properties of this flare based on imaging data from six AIA EUV passbands, including 131{\AA} (Fe XXI, $\sim$11 MK; Fe VIII, $\sim$0.4 MK), 94{\AA} (Fe XVIII, $\sim$7.1 MK; Fe X, $\sim$1.1 MK), 335{\AA} (Fe XVI, $\sim$2.5 MK), 211{\AA} (Fe XIV, $\sim$2.0 MK), 193{\AA} (Fe XII, $\sim$1.6 MK; Fe XXIV,  $\sim$17.8 MK), and 171{\AA} (Fe IX, $\sim$0.6 MK) \citep{ODwyer2010}. 
We adopt the method of \citet{Cheung2015}, who use a sparse inversion code to calculate the emission measure (EM) as a function of temperature from  AIA imaging data. This sparse inversion code is further updated by \citep{Su2018}, who adjusts the parameters of the sparse code to better suppress spurious high EM values at high temperatures.
Thus, the new differential emission measure (DEM, describing the amount of thermal plasma along the line of sight (LOS) as a function of T) diagnostic derived from the same AIA data is much more consistent with thermal X-ray observations.

In our calculation, 
we have re-binned the AIA images into a pixel size of 1.2$\arcsec$ (2$\times$2 rebinned) and used average intensities from two adjacent frames at 24s time cadence (a rebinning of 2$\times$ in time) for a better signal-to-noise ratio. The obtained EM is the line-of-sight integrated measure per unit area across the images. 

Here, we use the EM-weighted temperature $T_{\text{EM}}$ per pixel defined in the following formula \citep{Su2018} to construct the temperature map in spatial domain:
\begin{equation} 
T_{\text{EM}}=\frac{\sum{(\text{DEM}(T) \cdot \Delta T \cdot T)}}{\sum (\text{DEM}(T) \cdot \Delta T)}=\frac{\sum{(\text{EM}_{T} \cdot T)}}{\sum{\text{EM}_T}}
\end{equation}
From the temperature map (Figure~\ref{f3}(a)-(d) and the associated animation), one can clearly identify three episodes of the flare thermal evolution: In the first episode, a hot region stood out in the southernmost part of the active region, which corresponded to the group of sheared loops in Figure~\ref{f2}(a), later on, a propagation of thermal sources proceeded sequentially in space along the long arm of the $\Gamma$-shaped PIL during the second episode (Figure~\ref{f3}(b)), in the last episode, a hot arched region was newly formed, straddling the short arm of the $\Gamma$-shaped PIL (Figure~\ref{f3}(c) and (d)).\par
It is worth noting that, around 16:28 UT, there appeared two discrete thermal components both in the temperature map and the intensity contour of (thermally dominated) 6-25 keV RHESSI image (Figure~\ref{f3}(c)), these two components simultaneously existed in separated locations.
The thermal source with a higher projected position results from a continual migration along the extended curved PIL from the southeast to the northwest, while the one with a lower projected position was located at the top of a newly formed post-flare loop, as manifested by its apparent rise motion. The higher thermal source is relatively hotter than that of the lower thermal source. The centroid locations of these two components are separated by $\sim$ 70 $\arcsec$.\par

High-resolution HXR spectroscopy provides a powerful complement to imaging observations. 
The RHESSI spectrometer consists of an array of nine segmented germanium detectors (GeDs). 
Each detector is segmented into a thin front segment, which records photons from 3 keV to
2.7 MeV, with a resolution (FWHM) of  1 keV (at 100 keV), and a thick rear segment built to detect photons from about 20 keV to 17 MeV, with a resolution of  3 keV \citep[at 1 MeV,][]{Wigger2004}. 
Because of the strong attenuation below $\sim$ 6 keV and the K-escape events (The majority of the counts recorded below 6 keV is K-shell photon escaped from the germanium detector bombarded with high energy photons), no information can be gained about the incident photon spectrum below 6 keV\citep{Phillips2006}. Thus the energy fitting range is restricted above 6 keV.
The X-ray spectrum here is generated using the combined RHESSI front detectors \#1,3,5,6,8,9 to balance the resolution and signal-to-noise ratio (SNR). Among the excluded detectors, \#2 and \#7 show significantly worse energy resolution than the other detectors, the photon spectrum recorded in \#4 appears abnormal during this event.
The nonsolar background spectrum is selected during the neighboring RHESSI nighttime just before and/or just after the flare of interest. In order to ensure a reliable data set of RHESSI, care is taken to avoid the effect of attenuator state changes, satellite night times, South Atlantic Anomaly (SAA), and other complexities like photon pileup and decimation of data due to instrumental overflood.
For the spatially integrated spectra, we used the forward modeling method implemented by the Object Spectral Executive \citep[OSPEX;][]{Schwartz2002}. OSPEX uses an assumed parametric form
of the photon spectrum and finds parameter values that provide
the best fit in a $\chi^2$ value relating the observed background-subtracted photon flux with the predicted photon flux computed by folding the assumed incident photon spectrum through the spectrometer response matrix (DRM).

\begin{figure*} 
     \vspace{-0.0\textwidth}    
     \centerline{\hspace*{0.00\textwidth}
     \includegraphics[width=1.0\textwidth,clip=]{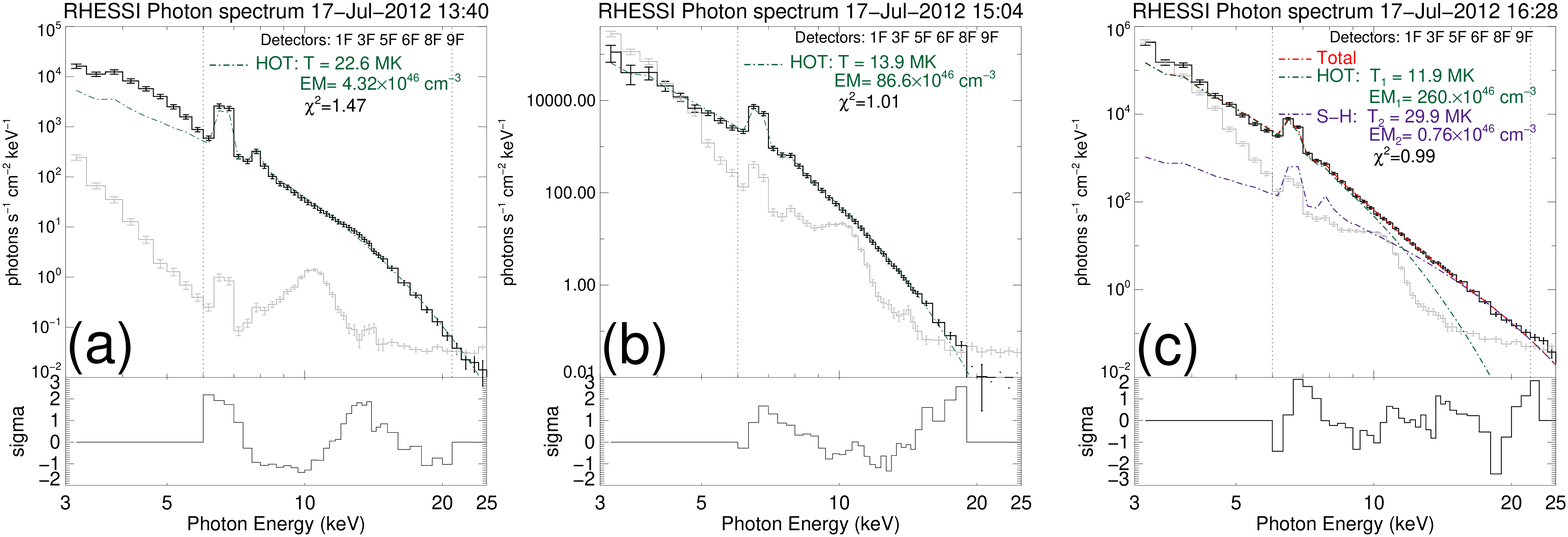}
               }
\caption{Panels (a)-(c): Photon flux spectra (black), model fit (hot component: green; super-hot: purple; total model: red), nonsolar background (grey), and normalized residuals during three peaks ($\sim$ 13:40, 15:05  and 16:28 UT), detectors 1F,3F,5F,6F,8F, and 9F are used}. \label{f4}
\end{figure*}

The thermal model (single temperature, $f\_vth$ in OSPEX) provides the plasma temperature T [\mbox{keV}] and volumetric emission measure EM [$\mbox{cm}^{-3}$] of the thermal source. The temperature and emission measure are free parameters while the relative iron abundance is fixed by default at the coronal value in the CHIANTI atomic database \citep{Dere1997,Landi2013}.
The direct evidence (HXR observation of footpoint source or emission above 20 keV) for high-energy energetic particles is absent during the flare impulsive phase (see Figure~\ref{f2}, accompanying animation, and Figure~\ref{f4}), which makes the nonthermal component fitting not considered.
Based on the thermal evolution of the flare (see Figure~\ref{f3}(a)-(d)), the spatially integrated spectra around 13:42 and 15:04 UT are fitted with a single-temperature thermal spectrum ($f\_vth$), while the spectrum around 16:28 UT is fitted with two isothermal functions ($f\_vth$+$f\_vth$). 
The fitting result of these intervals are presented in Figure~\ref{f4}(a)-(c).
It is found that the X-ray spectrum around 16:28 UT is well fitted by two distinct thermal components: a hot component ($\sim$11.9 MK) and a superhot component ($\sim$29.9 MK), yielding reduced $\chi^2$  value $\sim$ 0.99 (Figure~\ref{f4}(c)). \par

To identify precise locations of these two distinct thermal components,  we trace this flare evolution in the dual views of SDO AIA and STEREO-A EUVI with 120\mbox{\textdegree} separation angle (Figure~\ref{f5}(e) and the associated animation). The two thermal components are directly observed in the hot AIA 131 (10 MK) and/or AIA 94 (6.4 MK) passbands (Figure~\ref{f3}(e) and (f)), but are absent in cool AIA and EUVI passbands.  
Luckily,  by the
time they cool down to the STEREO EUVI 195\mbox{\AA} temperature range,
then the associated post-flare loop tops can be visible.
The AIA 193\mbox{\AA} filter has a similar response function to the STEREO/EUVI counterpart (193\mbox{\AA} $\rightarrow$ 195\mbox{\AA}).
We use a routine called ``scc\_measure.pro'' \citep{Thompson2009,Zhou2017} to determine the 3D structure 
from combined EUVI images from STEREO-A and SDO. 
The routine is a widget-based application that allows the user interactively to identify the same features in both images, then the 3D coordinates are calculated by the triangulation method.
Tops of their post-flare loop are marked in Figure~\ref{f5}(a) and (b), separately.  It is clear that there exist two distinct loop systems (Figure~\ref{f5}(b)) producing these two thermal components,
one is a sequence of flare loops stretching along the PIL from the south to the north, the other is a compact flare loop concentrating on a small region at the northmost part.


\section{Discussion and Conclusion} \label{sec:Conclusions}
\begin{figure*} 
      \vspace{-0.03\textwidth}    
      \centerline{\hspace*{0.00\textwidth}
      \includegraphics[width=1.0\textwidth,clip=]{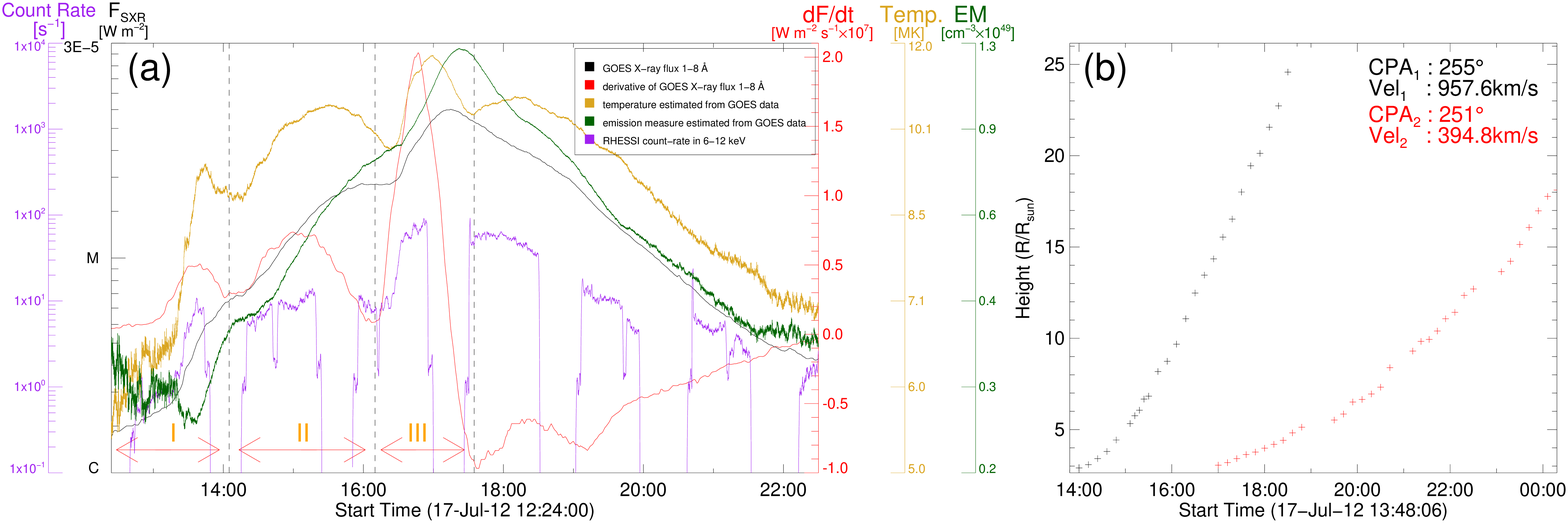}
      }
\caption{ (a) Temporal evolution of GOES soft X-ray flux (black), its time derivative (red), temperature (gold), emission measure (dark green), and RHESSI 6-12 keV count-rate (purple) of the observed M1.7 flare. According to the three peaks in the time derivative of GOES SXR (red profile), three episodes are marked by red segments in the lower-left of this panel; (b) The height-time plots for the leading edge of two flare-associated CMEs, the heights $R/R_{sun}$ (in solar radii with respect to the disk center) are measured at the fastest segment of the leading edge, the CPA, which is defined as the midangle of the two side edges of the CME in the sky plane, represents the location of the CME. The Velocity (Vel) gives the linear speed of the CMEs.} \label{f1}
\end{figure*}

\begin{figure*} 
      \vspace{-0.03\textwidth}    
      \centerline{\hspace*{0.00\textwidth}
      \includegraphics[width=1.0\textwidth,clip=]{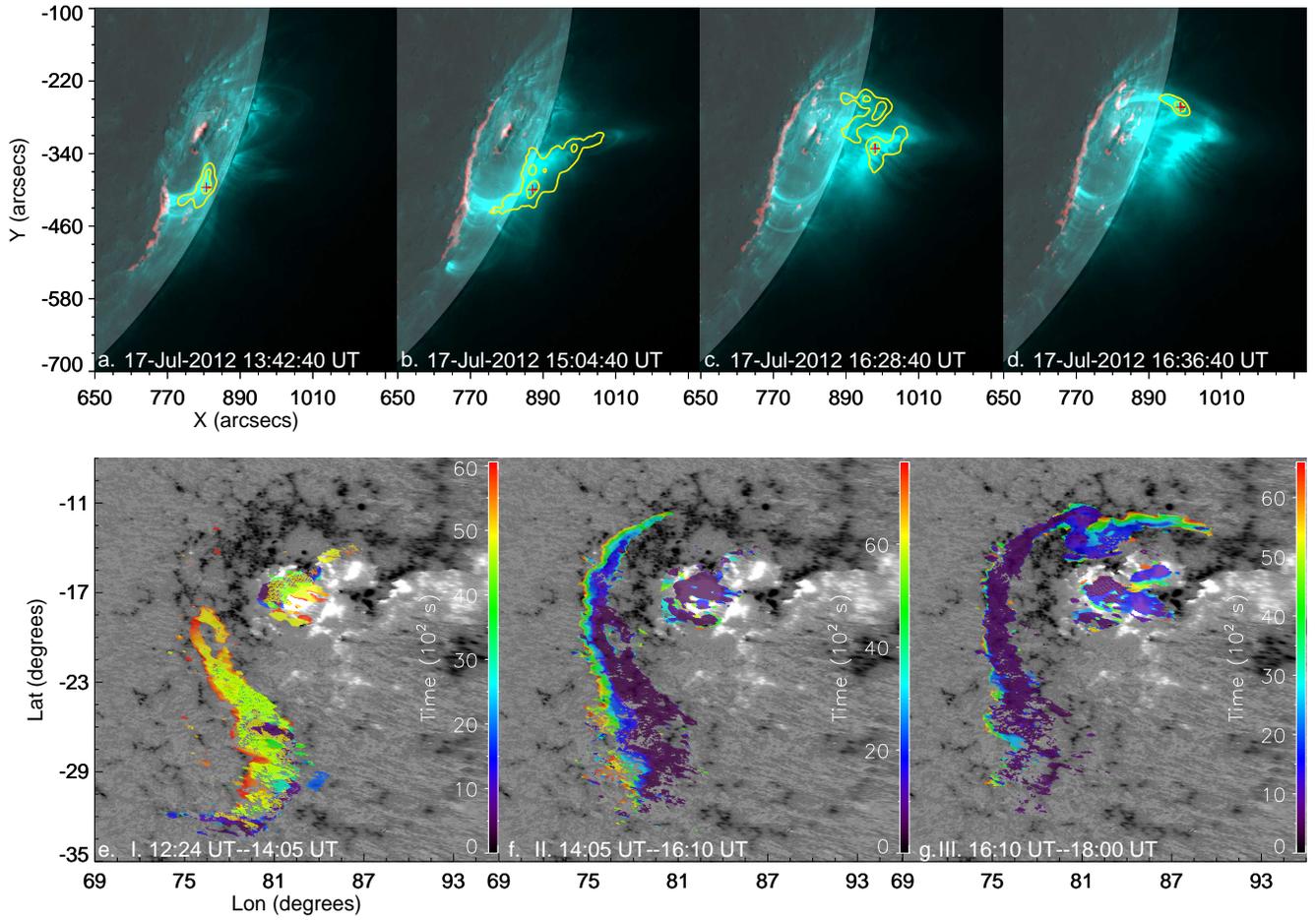}
      }
\caption{$Top:$ An overview of the evolution of the flare on July 17, 2012. (a)-(d) Composite images of AIA 131{\AA} (cyan) and 1600{\AA} (red) and HMI line-of-sight magnetic field (gray) at 13:42, 15:04, 16:28 and 16:36 UT, respectively. The overplotted contours in yellow indicate the emission of 6-25 keV from RHESSI at levels of 50\% and 80\% with its maximum marked by a plus symbol (red).  
An animation of these panels is available starting on 17 July 2012, 12:00:40 until 17 July 2012, 18:58:40 UT. The video duration is 17s. 
$Bottom:$ Evolution of the flare footpoint ribbons. The sequentially brightening of the flare ribbons are plotted with different colors from blue (the earliest time) to red (the lasted time). The background gray-scale image is the radial magnetic field of HMI magnetogram at 12:00 UT.
(Animation of this figure is available.)} \label{f2}
\end{figure*}

\begin{figure*} 
     \vspace{-0.0\textwidth}    
     \centerline{\hspace*{0.00\textwidth}
     \includegraphics[width=1.0\textwidth,clip=]{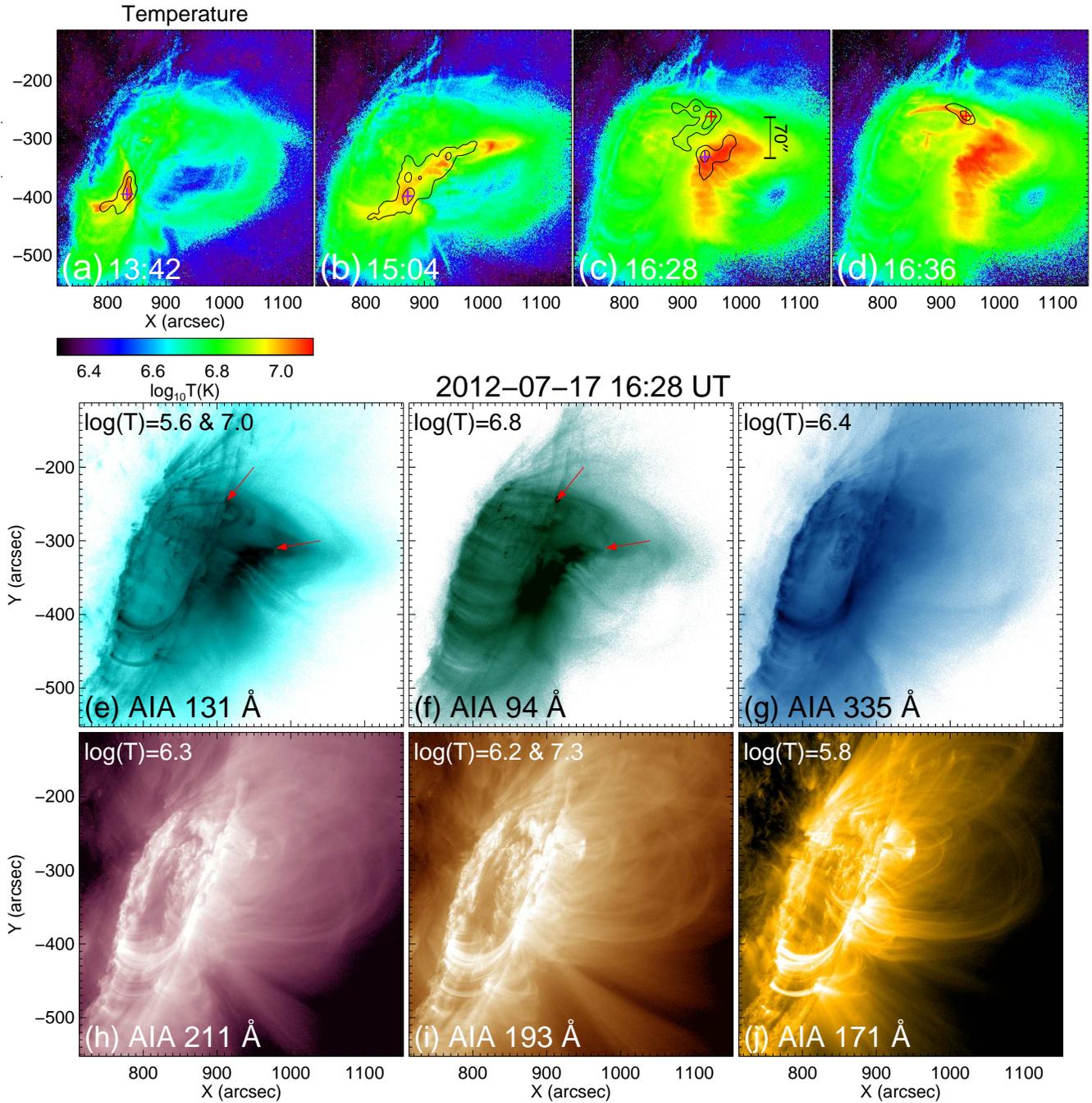}
               }
\caption{$Top:$ panels (a)-(d), the thermal evolution of the flare at 13:42, 15:04, 16:28, and 16:36 UT, RHESSI hard X-ray (6-25 keV, black) sources shown by the contours overlaid on the  temperature map, The contour levels are 50\% and 80\% of the peak flux; the plus symbols denote the derived centroid locations of the two distinct thermal components (purple and red). 
An animation of these panels is available starting on 17 July 2012, 12:00:20 until 17 July 2012, 16:59:32 UT. The video duration is 31 s. 
$Bottom:$ SDO/AIA observations from each of the six coronal filters during 16:28 UT of the 17-Jul-2012 flare, panels (e)-(g) are reverse color images, the hot features in (e) and (f) are identified by the red arrows which are absent in the other AIA channels.  (Animation of this figure is available.)} \label{f3}
\end{figure*}

Imaging and spectroscopic observations of this event show that the superhot plasma ($\sim$30 MK) is distinct, both spectrally and spatially, from the usual $\sim$ 10--20 MK plasma.  The flare consists of two separate loop structures at different locations. The longer loop system develops with persistently higher temperatures, after the second CME; the shorter loop system dominates mainly the SXR emission. 
In the case of \citet{Caspi2010}, the centroids of the spatially distinct superhot and hot sources are separated by 10$\arcsec$. In our observations, their projected distance is 70$\arcsec$. Also, 
the three episodes of HXR enhancements as proxied by the time derivative of the SXR in Fig.~\ref{f1}(a) are associated with two CMEs, which implies that the whole event went through three successive flare processes separately, rather than an extended heating process. 
Imaging observations in Fig.~\ref{f2} corroborate the general pattern of the coronal loop and footpoint evolution, consistently confirming the three enhancements of HXR emission.  The superhot component in Fig.~\ref{f3}(b) and (c) is the extended structure originating from the second stage evolution. The hot component in Fig.~\ref{f3}(c) and (d), however, originated from the top of the post flare loop emerging in the last stage evolution.

To summarize the observations, the overall evolution of the two thermal components, including their actual location of the creation site, and the timing and relationship to the flare-energy release, is  summarized as follows:

Accompanied with the first CME, the apparent sequential evolution along the PIL is attributed to numerous episodes of similar magnetic reconnection occurring successively along the long arm of the $\Gamma$-shaped PIL. Magnetic reconnection commences in the southernmost part of the active region resulting in a superhot region beneath. The thermal energy is then transported along the magnetic field lines toward the footpoint of the magnetic loops via thermal conduction, producing the flare ribbon and chromospheric evaporation toward the loop top. The evaporated plasma is supposed to be further heated at the loop-top region, contributing to the increase of the emission measure of the hot plasma there. Later on, the other distinct loop system appears along the short arm of the $\Gamma$-shaped PIL and produces the second CME. Initiated by the magnetic reconnection, similar energy release and transport process occur in this set of loop system. Compared with the first flare loop system, this set of loop system is compact and its projection height is lower.

\begin{figure*} 
     \vspace{-0.0\textwidth}    
     \centerline{\hspace*{0.00\textwidth}
     \includegraphics[width=1.0\textwidth,clip=]{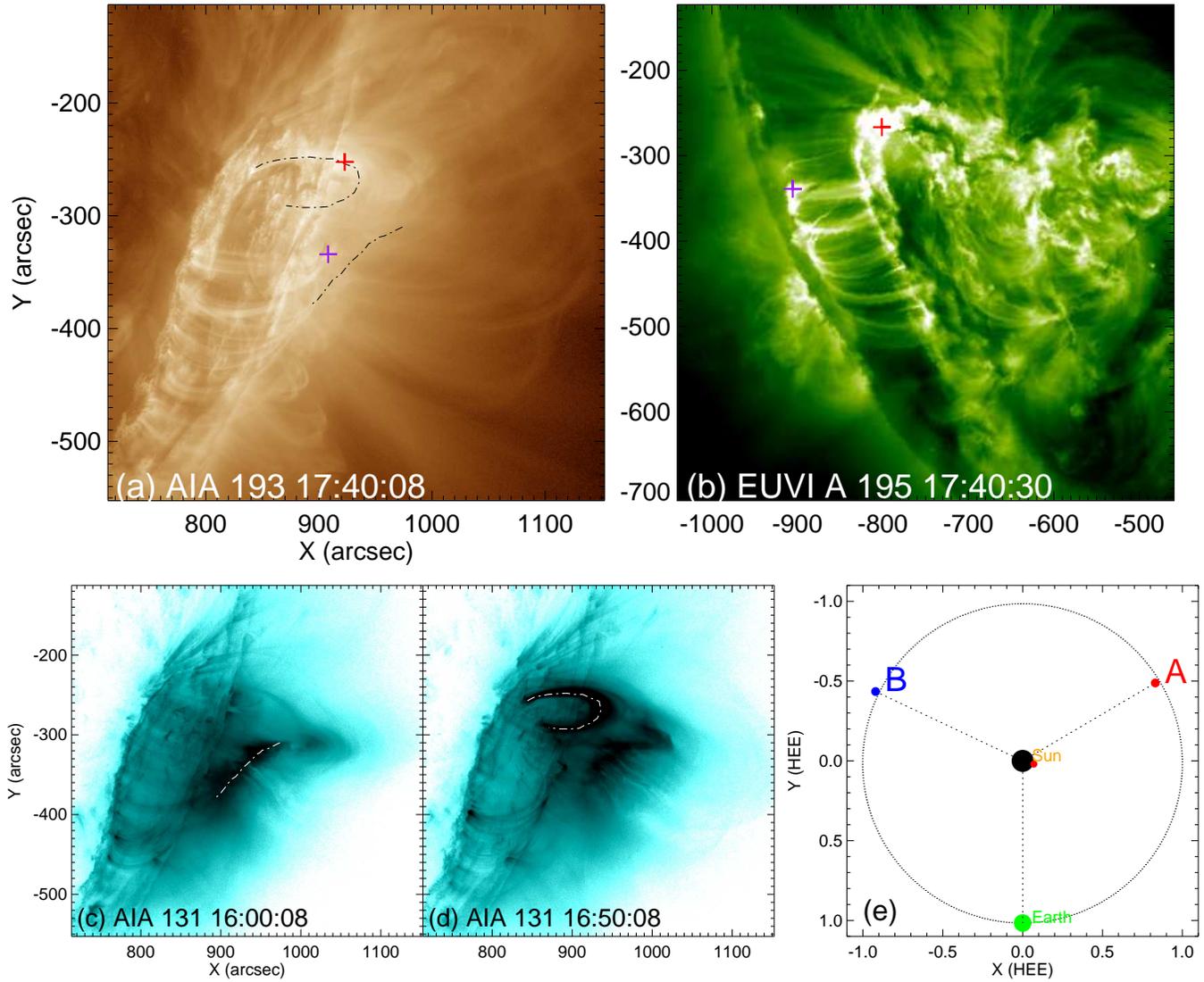}
               }
\caption{$Top:$ SDO/AIA 193$\mbox{\AA}$ and STEREO-A/EUVI 195$\mbox{\AA}$ images at about 17:40 UT during the eruption. The plus symbols denote the post flare loop tops used for 3D triangulation. The black dotted dash lines depict the location of the hot components,
An animation of these panels is available starting on 17 July 2012, 12:00 until 17 July 2012, 18:59 UT. The video duration is 1m24s. 
$Bottom:$ The dotted dash lines in panels (c) and (d) show the extended flare loop system at 16:00 UT and the compact flare loop system at
16:50 UT; Panel (e): Positions of the STEREO-A/B and Earth (SDO) in the ecliptic plane on 2012 July 17. The red dot on the Sun marks the flare source region, which appears on the solar disk when viewed from STEREO-A, on the limb from SDO, and on the backside of the Sun from STEREO-B. (Animation of this figure is available.)} \label{f5}
\end{figure*}

From a side view including the Earth perspective (e.g., Figure~\ref{f2}(c)), the superhot component is located above the flare loop top in projection, similar to the cases reported by \citet{Svestka1985, Caspi2010}. But from the top view (e.g., Figure~\ref{f5}(b)), the two distinct thermal components are associated with two separate loop structures with the longer one having higher temperatures, which is reminiscent of the cases reported by \citet{Den1989,Nitta1997}. It is not rare for homologous eruptions to originate from different segments of the same extended PIL within a short time interval \citep[e.g.,][]{Shen2013,Liu2017}. With different flare loop systems cooling with different timescales and with the flare maximal temperature correlated with the flare class \citep{Caspi2014}, it is anticipated that flare plasmas in a set of homologous flares may be manifested as two or more co-existing thermal components of different temperatures in HXRs, which are most likely distinct in space, but in certain circumstances may also appear co-spatial in projection \citep[e.g.][]{Sharykin2015}.

In fact, this unique event has the longest impulsive phase of any M or X-class flare in the present Hale cycle since 1995,  based on a search of the GOES database\footnote[1]{http://sprg.ssl.berkeley.edu/$\sim$tohban/wiki/index.php/The\underline{ }Slowest\underline{ }Flare}. The slow and large-scale evolution provides an excellent opportunity to perform an intricate structure of the thermal source locations and study their evolutions. 
And, thanks to multiperspective observations, this is the first time attempting to resolve the locations of the hot and superhot sources in 3D, revealing the two sources coming from two different loop systems instead of different altitudes of the same loop system.

\acknowledgements The authors wish to express their special thanks to the referee for suggestions
and comments which led to the improvement of the paper. The authors thank S\"{a}m Kruker, Wei, Liu, and Brian R. Dennis for constructive help on RHESSI data analysis and the teams of \emph{SDO}, \emph{RHESSI}, \emph{STEREO} and \emph{GOES} for excellent data set.

This work is supported by the B-type Strategic Priority Program XDB41000000 funded by the Chinese
Academy of Sciences.
Z.J. is supported by NSFC grants 42004142 and LDSE201703 from Key Laboratory of Lunar and Deep Space Exploration, CAS. 
R.L. and Y.W. are supported by NSFC grants 41574165, 41761134088, 41774150, 11925302, and 41774178.
J.S., M.D., and X.C. are supported by NSFC grants 11722325, 11733003, 11790303, 11790300, and Jiangsu NSF grant BK20170011.
 L.L. is supported by NSFC grant 11803096. 
J.C. acknowledges support by NSFC grants 41822404, 41731067, 41574170, 41531073.

\begin{thebibliography}{50}
\expandafter\ifx\csname natexlab\endcsname\relax\def\natexlab#1{#1}\fi

\bibitem[Allred et al.(2005)]{Allred2005} Allred, J.~C., Hawley, S.~L., Abbett, W.~P., et al.\ 2005, \apj, 630, 573

\bibitem[Allred et al.(2015)]{Allred2015} Allred, J.~C., Kowalski, A.~F., \& Carlsson, M.\ 2015, \apj, 809, 104

\bibitem[Antiochos \& Sturrock(1978)]{Antiochos1978} Antiochos, S.~K., \& Sturrock, P.~A.\ 1978, \apj, 220, 1137 

\bibitem[Antonucci et al.(1984)]{Antonucci1984} Antonucci, E., Gabriel, A.~H., \& Dennis, B.~R.\ 1984, \apj, 287, 917. doi:10.1086/162749



\bibitem[Brown(1971)]{Brown1971} Brown, J.~C.\ 1971, \solphys, 18, 489. doi:10.1007/BF00149070


\bibitem[{{Caspi} {et~al.}(2014){Caspi}, {Krucker}, \& {Lin}}]{Caspi2014}
{Caspi}, A., {Krucker}, S., \& {Lin}, R.~P. 2014, \apj, 781, 43

\bibitem[{{Caspi} \& {Lin}(2010)}]{Caspi2010}
{Caspi}, A., \& {Lin}, R.~P. 2010, \apjl, 725, L161

\bibitem[Cheng et al.(2019)]{Cheng2019} Cheng, Z., Wang, Y., Liu, R., Zhou, Z., \& Liu, K.\ 2019, \apj, 875, 93
\bibitem[Cheung et al.(2015)]{Cheung2015} Cheung, M.~C.~M., Boerner, P., Schrijver, C.~J., et al.\ 2015, \apj, 807, 143

\bibitem[Dhakal et al.(2018)]{Dhakal2018} Dhakal, S.~K., Chintzoglou, G., \& Zhang, J.\ 2018, \apj, 860, 35. doi:10.3847/1538-4357/aac028

\bibitem[Den \& Somov(1989)]{Den1989} Den, O.~G. \& Somov, B.~V.\ 1989, \sovast, 33, 149


\bibitem[Dennis \& Zarro(1993)]{Dennis1993} Dennis, B.~R. \& Zarro, D.~M.\ 1993, \solphys, 146, 177. doi:10.1007/BF00662178
\bibitem[Dere et al.(1997)]{Dere1997} Dere, K.~P., Landi, E., Mason, H.~E., et al.\ 1997, \aaps, 125, 149

\bibitem[Fisher et al.(1985)]{Fisher1985} Fisher, G.~H., Canfield, R.~C., \& McClymont, A.~N.\ 1985, \apj, 289, 425. doi:10.1086/162902

\bibitem[Howard et al.(2008)]{Howard2008} Howard, R.~A., Moses, J.~D., Vourlidas, A., et al.\ 2008, \ssr, 136, 67. doi:10.1007/s11214-008-9341-4

\bibitem[Hurford et al.(2002)]{Hurford2002} Hurford, G.~J., Schmahl, E.~J., Schwartz, R.~A., et al.\ 2002, \solphys, 210, 61. doi:10.1023/A:1022436213688


\bibitem[Kaiser et al.(2008)]{Kaiser2008} Kaiser, M.~L., Kucera, T.~A., Davila, J.~M., et al.\ 2008, \ssr, 136, 5. doi:10.1007/s11214-007-9277-0


\bibitem[Korneev et al.(1979)]{Korneev1979} Korneev, V.~V., Krutov, V.~V., Mandelshtam, S.~L., et al.\ 1979, \solphys, 63, 319. doi:10.1007/BF00174537

\bibitem[Landi et al.(2013)]{Landi2013} Landi, E., Young, P.~R., Dere, K.~P., et al.\ 2013, \apj, 763, 86

\bibitem[Lemen et al.(2012)]{Lemen2012} Lemen, J.~R., Title, A.~M., Akin, D.~J., et al.\ 2012, \solphys, 275, 17. doi:10.1007/s11207-011-9776-8


\bibitem[Lin et al.(1981)]{Lin1981} Lin, R.~P., Schwartz, R.~A., Pelling, R.~M., et al.\ 1981, \apjl, 251, L109. doi:10.1086/183704

\bibitem[Lin et al.(2002)]{Lin2002} Lin, R.~P., Dennis, B.~R., Hurford, G.~J., et al.\ 2002, \solphys, 210, 3. doi:10.1023/A:1022428818870

\bibitem[{{Liu} {et~al.}(2013){Liu}, {Chen}, \& {Petrosian}}]{Liu2013}
{Liu}, W., {Chen}, Q., \& {Petrosian}, V. 2013, \apj, 767, 168
\bibitem[Liu et al.(2017)]{Liu2017} Liu, L., Wang, Y., Liu, R., et al.\ 2017, \apj, 844, 141. doi:10.3847/1538-4357/aa7d56

\bibitem[Nitta \& Yaji(1997)]{Nitta1997} Nitta, N. \& Yaji, K.\ 1997, \apj, 484, 927. doi:10.1086/304360

\bibitem[{{O'Dwyer} {et~al.}(2010){O'Dwyer}, {Del Zanna}, {Mason}, {Weber}, \&
  {Tripathi}}]{ODwyer2010}
{O'Dwyer}, B., {Del Zanna}, G., {Mason}, H.~E., {Weber}, M.~A., \& {Tripathi},
  D. 2010, \aap, 521, A21

  \bibitem[Ogawara et al.(1991)]{Ogawara1991} Ogawara, Y., Takano, T., Kato, T., et al.\ 1991, \solphys, 136, 1. doi:10.1007/BF00151692


\bibitem[{{Pesnell} {et~al.}(2012){Pesnell}, {Thompson}, \&
  {Chamberlin}}]{Pesnell2012}
{Pesnell}, W.~D., {Thompson}, B.~J., \& {Chamberlin}, P.~C. 2012, \solphys,
  275, 3
\bibitem[Pike et al.(1996)]{Pike1996} Pike, C.~D., Phillips, K.~J.~H., Lang, J., et al.\ 1996, \apj, 464, 487. doi:10.1086/177338
  
\bibitem[Schou et al.(2012)]{Schou2012} Schou, J., Scherrer, P.~H., Bush, R.~I., et al.\ 2012, \solphys, 275, 229. doi:10.1007/s11207-011-9842-2

  
\bibitem[Schwartz et al.(2002)]{Schwartz2002} Schwartz, R.~A., Csillaghy, A., Tolbert, A.~K., et al.\ 2002, \solphys, 210, 165
\bibitem[Sharykin et al.(2015)]{Sharykin2015} Sharykin, I.~N., Struminskii, A.~B., \& Zimovets, I.~V.\ 2015, Astronomy Letters, 41, 53. doi:10.1134/S1063773715020061

\bibitem[Sheeley et al.(1983)]{Sheeley1983} Sheeley, N.~R., Howard, R.~A., Koomen, M.~J., et al.\ 1983, \apj, 272, 349. doi:10.1086/161298


\bibitem[Shen et al.(2013)]{Shen2013} Shen, C., Li, G., Kong, X., et al.\ 2013, \apj, 763, 114. doi:10.1088/0004-637X/763/2/114

\bibitem[Smith et al.(2002)]{Smith2002} Smith, D.~M., Lin, R.~P., Turin, P., et al.\ 2002, \solphys, 210, 33. doi:10.1023/A:1022400716414


\bibitem[Su et al.(2018)]{Su2018} Su, Y., Veronig, A.~M., Hannah, I.~G., et al.\ 2018, \apjl, 856, L17

\bibitem[{{Sun} {et~al.}(2014){Sun}, {Cheng}, \& {Ding}}]{Sun2014}
{Sun}, J.~Q., {Cheng}, X., \& {Ding}, M.~D. 2014, \apj, 786, 73

\bibitem[{{Sun} {et~al.}(2016){Sun}, {Zhang}, {Yang}, {Cheng}, \&
  {Ding}}]{Sun2016}
{Sun}, J.~Q., {Zhang}, J., {Yang}, K., {Cheng}, X., \& {Ding}, M.~D. 2016,
  \apjl, 830, L4


\bibitem[Svestka \& Poletto(1985)]{Svestka1985} Svestka, Z. \& Poletto, G.\ 1985, \solphys, 97, 113. doi:10.1007/BF00152982


\bibitem[Tanaka(1987)]{Tanaka1987} Tanaka, K.\ 1987, \pasj, 39, 1

\bibitem[Thompson(2009)]{Thompson2009} Thompson, W.~T.\ 2009, \icarus, 200, 351. doi:10.1016/j.icarus.2008.12.011

\bibitem[Phillips et al.(2006)]{Phillips2006} Phillips, K.~J.~H., Chifor, C., \& Dennis, B.~R.\ 2006, \apj, 647, 1480. doi:10.1086/505518


\bibitem[Webb \& Hundhausen(1987)]{Webb1987} Webb, D.~F. \& Hundhausen, A.~J.\ 1987, \solphys, 108, 383. doi:10.1007/BF00214170


\bibitem[Wigger et al.(2004)]{Wigger2004} Wigger, C., Hajdas, W., Arzner, K., et al.\ 2004, \apj, 613, 1088

\bibitem[Woodgate et al.(1984)]{Woodgate1984} Woodgate, B.~E., Martres, M.-J., Smith, J.~B., et al.\ 1984, Advances in Space Research, 4, 11. doi:10.1016/0273-1177(84)90151-0

\bibitem[Zhou et al.(2017)]{Zhou2017} Zhou, Z., Zhang, J., Wang, Y., et al.\ 2017, \apj, 851, 133. doi:10.3847/1538-4357/aa9bd9


\end{thebibliography}

\end{CJK*}
\end{document}